\documentclass[journal]{IEEEtran}

\ifCLASSINFOpdf
   \usepackage[pdftex]{graphicx}
   \usepackage{subfigure}
   \DeclareGraphicsExtensions{.pdf,.jpeg,.png}
\else
 
\fi

\usepackage{adjustbox}
\usepackage{caption}
\usepackage{amsmath}
\usepackage{amssymb}
\usepackage{color}
\usepackage{multirow}
\begin{document}

\title{Attention-Aware Linear Depthwise Convolution\\ for Single Image Super-Resolution}
%
%
%

\author{Seongmin Hwang, Gwanghuyn Yu,
        Cheolkon Jung,~\IEEEmembership{Member,~IEEE,}
        and~Jinyoung Kim
        
\thanks{This work was supported by the Ministry of Science and ICT (MSIT), Korea, under the High-Potential Individuals Global Training Program) (2019-0-01609) supervised by the Institute for Information Communications Technology Planning Evaluation (IITP). Also, this work was supported by the National Natural Science Foundation of China (No. 61872280) and the International S\&T Cooperation Program of China (No. 2014DFG12780).}        
\thanks{
    S. Hwang, G. Yu, and J. Kim are with the School of Electronic and Computer Engineering, Chonnam National University, Yongbong-ro 77, Gwangju 61186, South Korea e-mail: 197209@jnu.ac.kr; sayney1004@naver.com; beyondi@chonnam.ac.kr.
    
    C. Jung is with the School of Electronic Engineering, Xidian University, Xi'an 710071, China e-mail: zhengzk@xidian.edu.cn.

(Corresponding authors: Cheolkon Jung and Jinyoung Kim).}
\thanks{}}

\markboth{}%
{Shell \MakeLowercase{\textit{et al.}}: Bare Demo of IEEEtran.cls for IEEE Journals}
%

\maketitle

\begin{abstract}
Although deep convolutional neural networks (CNNs) have obtained outstanding performance in image super-resolution (SR), their computational cost increases geometrically as CNN models get deeper and wider. Meanwhile, the features of intermediate layers are treated equally across the channel, thus hindering the representational capability of CNNs. In this paper, we propose an attention-aware linear depthwise network to address the problems for single image SR, named ALDNet. Specifically, linear depthwise convolution allows CNN-based SR models to preserve useful information for reconstructing a super-resolved image while reducing computational burden. Furthermore, we design an attention-aware branch that enhances the representation ability of depthwise convolution layers by making full use of depthwise filter interdependency. Experiments on publicly available benchmark datasets show that ALDNet achieves superior performance to traditional depthwise separable convolutions in terms of quantitative measurements and visual quality.
\end{abstract}


\IEEEpeerreviewmaketitle

\section{Introduction}

\IEEEPARstart{S}{ingle} image super-resolution (SISR) has gained much attention for the past decade. In general, the SISR task aims at restoring high-resolution (HR) image from its corresponding low-resolution input. However, SISR inherently has an ill-posed nature since a number of HR images can take the same low-resolution (LR) image by down-sampling, the solution space for SR problems is extremely large and there exist multiple HR solutions for LR input which makes SR operation an one-to-many mapping from LR image to HR image. To solve the ill-posed problem, a number of SR methods have been proposed for decades to produce plausible reconstructions by super-resolving high-frequency information from a given LR input, ranging from interpolation-based \cite{interpolation} to learning based methods \cite{SRCNN, VDSR, EDSR, RDN}.

Among them, Dong et al. proposed SRCNN \cite{SRCNN} and firstly demonstrated that CNN could be used to learn a mapping LR image to HR image showing superior performance to traditional methods. Kim et al. \cite{VDSR} built a very deep network for SR (VDSR) using 20 layers by cascading small filters many times. Zhang et al. proposed the Residual Dense Network \cite{RDN} (RDN) which fully used hierarchical features from all the convolutional layers by combining residual skip connections with dense connections.

Despite learning-based models, especially CNN-based models have achieved significant advances in image SR, and they are still facing two main problems: (1) CNN-based SR models suffer from computational complexity and memory consumption in practice as CNN models get deeper and wider to learn more discriminative high-level features. To overcome this problem, Howard et al. proposed depthwise separable convolution \cite{mobilenet} that factorized a standard convolution into two  convolutions: (1) depthwise convolution as a spatial filter, (2) pointwise convolution as a lightweight feature generator. Although depthwise separable convolution successfully achieves its superiority especially in terms of time and memory consumption, it is designed to solve high-level vision tasks such as image classification and object detection. Thus, depthwise separable convolution should be optimally adjusted prior to being applied to the low-level vision task. (2) Traditional CNN-based models usually adopt cascade network topologies. However, in this way the features in intermediate layers are all treated equally without considering inherent feature correlations. Hence, the features of each layer are sent to the sequential layer without any distinction, thereby hampering discriminative ability of CNN-based models.

To mitigate these drawbacks above, we propose an attention-aware linear depthwsie network, named ALDNet, to preserve informative components as well as allow CNN-based SR models to have more powerful discriminative ability. The main idea of ALDNet is to remove non-linearity between depthwise convolution and pointwise convolution, thus making depthwise convolution linear mapping function. Linear depthwise convolution prevents non-linearity from destroying informative features, and thus helps reconstruct SR image. Furthermore, we build an attention-aware branch for depthwise convolution layer to adaptively recalibrate each channel-wise feature by modeling the interdependencies across all depthwise convolution filters. Such attention-aware branch makes full use of the advantage of depthwise convolution filter enabling ALDNet to strengthen more informative features. Hence, ALDNet is able to deal with the main problems that most of CNN-based models are facing. Experimental results demonstrate the superiority of ALDNet.

Compared with existing SR methods, the main contributions of this paper are summarized as follows:

\begin{itemize}
    \item We propose a novel ALDNet that learns correlations of depthwise convolution filters and remarkably enhances the discriminative capability of CNN models. To the best of our knowledge, this is the first work of using interdependencies among convolution filters to get attention.
    
    \item We present linear depthwise convolution to preserve informative features by removing non-linearity between depthwise convolution and pointwise convolution. 
    
    \item We build a baseline network for SISR based on attention-aware linear depthwise blocks, named ALDSR. ALDSR, which is a very lightweight network, shows superior performance in SR to other state-of-the-art SR models.
\end{itemize}

 \begin{figure}[t]
    \centering
    \subfigure[Depthwise separable convolution (existing)]{
    \includegraphics[page = {1}, scale=0.3]{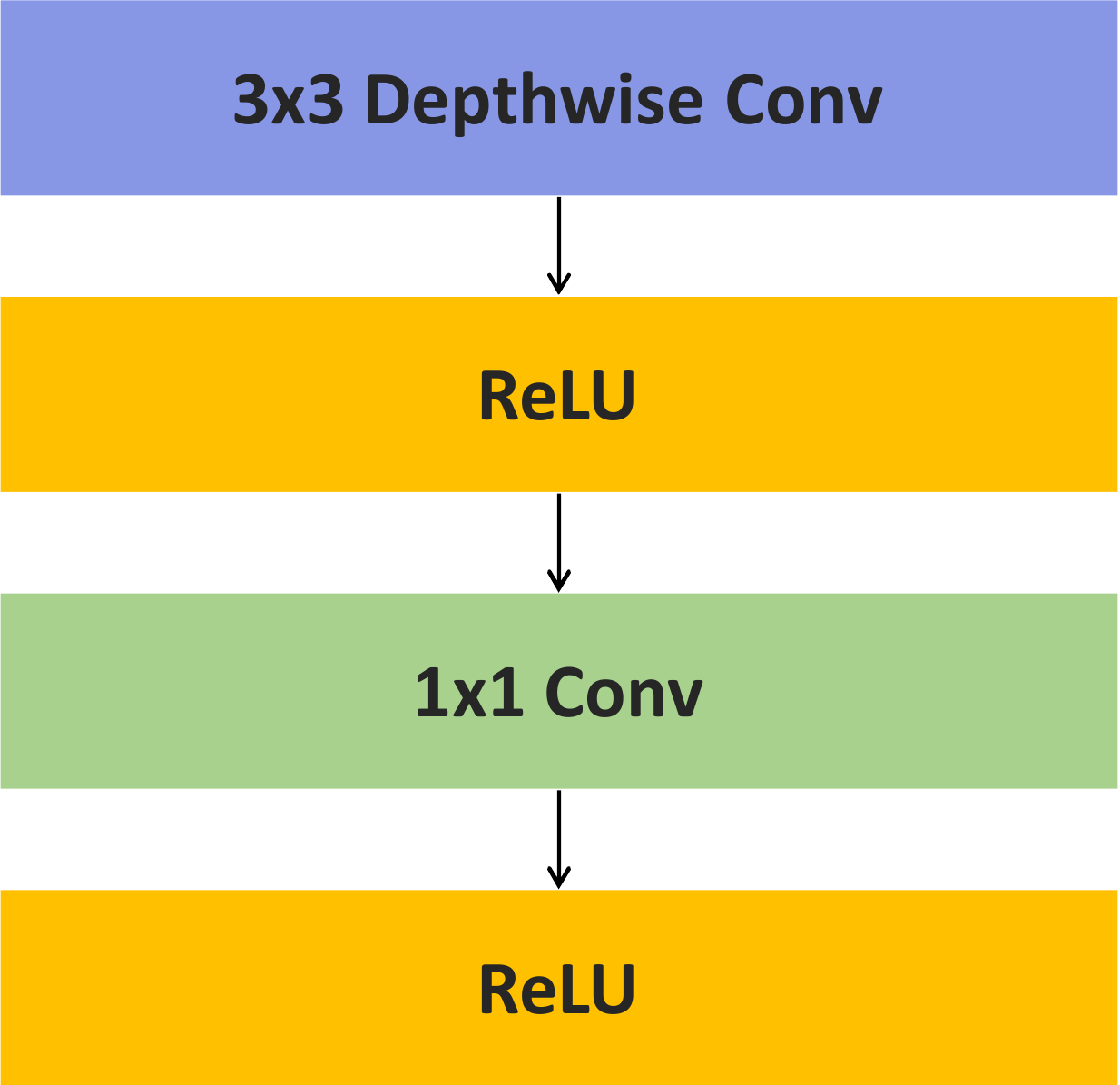}
    \label{fig1_a:DW}
    }
    \quad
    \centering
    \subfigure[Linear depthwise separable convolution (proposed)]{
    \includegraphics[page = {2}, scale=0.3]{01DWandLDW.pdf}
    \label{fig1_b:LDW}
    }
    \caption{Comparison between depthwise separable Convolution and linear depthwise separable convolution} 
    \label{fig1:DWandLWD}
 \end{figure}

\section{Attention-Aware Linear Depthwise Convolution}

\subsection{Linear Depthwise Convolution}
Recently, MobileNet \cite{mobilenet} proposed depthwise separable convolution and achieved outstanding performance in computer vision problems with low computational cost. However, depthwise separable convolution aimed to deal with high-level vision problems such as image classification. Thus, it is improper to apply depthwise separable convolution to low-level vision tasks such as image denoising and SR. Therefore, the MobileNet architecture should be modified to be used for image SR where the information should be handled more carefully. According to a former study \cite{EDSR} as shown in Fig. \ref{fig1_a:DW}, we illustrated depthwise separable convoltuion by removing batch normalization \cite{batchnorm} layers from standard depthwise separable convolution. Different from high-level vision tasks, since input and output domains of SR task are both images, the change in the distribution of network activations, so-called internal covariance shift \cite{batchnorm}, dose not occur severely during training. Instead, batch normalization would rather get rid of network flexibility and lose scale information of an image while increasing GPU memory consumption during the training process. In this sense, batch normalization is no longer efficient for SR problem.

In this work, we propose linear depthwise convolution in which non-linearity (ReLU) is eliminated between depthwise convolutional layer and pointwise convolutional layer as shown in Fig. \ref{fig1_b:LDW}. Depthwise separable convolution itself factorizes standard convolution by separating spatial filtering from the feature generation mechanism. Depthwise convolution, which acts as spatial filtering, performs convolution independently for every input channel with only considering spatial information.  1$\times$1 convolution, called pointwise convolution, computes linear combination considering channel information produced by depthwise convolution. That is, the feature map generated by depthwise convolution only considers spatial information. 

We denote ReLU between depthwise convolution and pointwise convolution as "hasty ReLU". Consequentially, "hasty ReLU" is utilized by only considering spatial information without any consideration of channel features. Thus, "hasty ReLU" could cause destruction of information in SR reconstruction. Since the information preservation is a crucial factor for image SR, it is necessary to have special care on using non-linearity such as ReLU that causes information loss. Although "hasty ReLU" might work well on high-level vision task, it is obviously not good for solving low-level vision problems. As a result, we do not adopt "hasty ReLU" to achieve significant SR performance improvement.

\subsection{Attention-Aware Depthwise Convolution}
Most previous CNN-based models generally treat the features in intermediate layers equally. To address this problem, SENet \cite{senet} was introduced to recalibrate the channelwise feature values in CNN-based models. In addition, recent works \cite{rcan, san} have shown that attention mechanism is helpful for improving SR performance to enhance more discriminative capability of CNNs. Typically, most of attention based CNN models adopt self-attention mechanism to get attention by capturing correlation between one pixel and other ones of the feature maps. 

 \begin{figure}[t]
 \centering
    \includegraphics[scale=0.3]{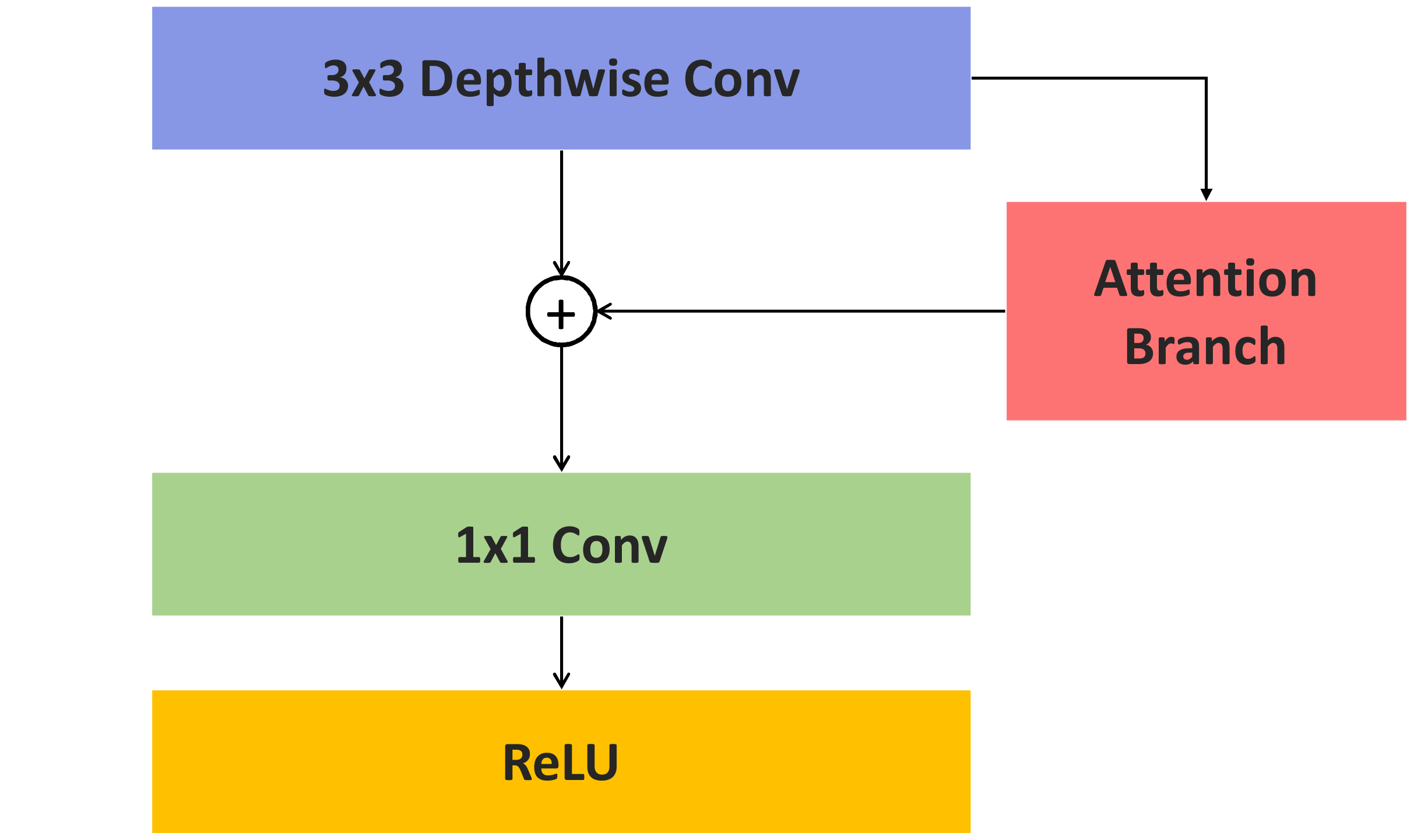}
    \captionsetup{justification=centering}
    \caption{Attention-aware linear depthwise convolution.} 
    \label{fig2:ALD}
 \end{figure}
 
However, in CNN, of course, feature maps are important, but convolutional layers which directly generate feature values also play vital role. In this sense, we could consider utilizing parameters of convolutional filters for attention mechanism. Furthermore, since depthwise convolution filter has very few parameters compared with standard convolution, it is easier and more effective to extract representation values for getting attention. Added to this, depthwise convolution produces current feature maps by applying a single filter to each input channel unlike standard convolution, hence, the current feature maps are incomplete. However, if the network learns the importance of each depthwise filter and recalibrates the feature values by correlation of depthwise filters, the quality of the feature maps are improved, thus leading to a remarkable improvement of the network performance.

Driven by the observations, we propose attention-aware depthwise convolution which enables a network to learn importance of each depthwise filter of depthwise convolutional layer. By fully exploiting the interdependencies among depthwise filters, we improve representation power of CNN models significantly. We build a channel attention branch to model depthwise convolution interdependencies and recalibrate channelwise features as follows. The architecture of attention-aware linear depthwise convolution, which incorporates an attention-aware branch to linear depthwise convolution, is illustrated in Fig. \ref{fig2:ALD}.
\vspace{0.2cm}

\textbf{Determinant of depthwise filter:}
Different from standard convolution, depthwise convolution has only single filter to produce feature maps, hence depthwise convolutional layer could be represented as a group of square matrices with the same number of rows and columns. Thus, we make full use of the distinctive feature of depthwise convolution as an informative descriptor. There are many methods to describe characteristics of square matrix. Among them, determinant captures important information about the matrix in a single number, and can be viewed as the volume scaling factor of the linear transformation by the square matrix. Since average and max pooling tend to oversimplify the filter information, we adopt determinant as the depthwise convolution descriptor.

 \begin{figure}[t]
 \centering
    \includegraphics[scale=0.5]{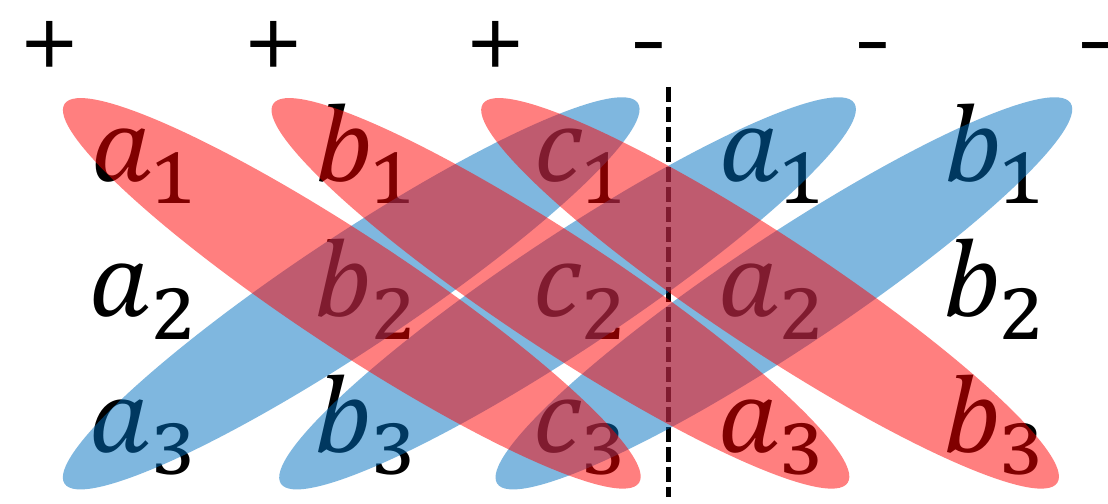}
    \captionsetup{justification=centering}
    \caption{Sarrus' rule to compute determinant.} 
    \label{fig4:sarrus}
 \end{figure}

Formally, given a group of depthwise filter $k\times k\times C$, $\mathbf{D} = [\mathbf{d}_{1}, \mathbf{d}_{2}, \ldots ,\mathbf{d}_{c}]$ with $C$ filters with kernel size of $k$. Each filter $\mathbf{d}_{1}, \mathbf{d}_{2}, \ldots ,\mathbf{d}_{c}$ can be represented as $k\times k$ square matrix as follows:
\begin{align}
W_{k}^{c} = 
\begin{bmatrix}
w_{1,1}^{c} & w_{1,2}^{c} & \cdots & w_{1,k}^{c} \\
w_{2,1}^{c} & w_{2,2}^{c} & \cdots & w_{2,k}^{c} \\
\vdots & \vdots & \ddots & \vdots \\
w_{k,1}^{c} & w_{k,2}^{c} & \cdots & w_{k,k}^{c}
\end{bmatrix}
\end{align}
where $w_{i,j}^{c}$ is a weight of convolution filter at position ($i, j$) of $c$-th channel.

In \cite{vggnet}, Simonyan et al. reported that cascading 3$\times$3 filters has the same effect as the use of large filter sizes such as 7$\times$7 and 11$\times$11, while reducing computational complexity. Thus, most of recent CNN-models adopt a combination of several convolutional layers with kernel size 3$\times$3. The corresponding depthwise convolutional filter of a standard 3$\times$3 convolutional filter is represented by
\begin{align}
W_{3}^{c} = 
\begin{bmatrix}
w_{1,1}^{c} & w_{1,2}^{c} & w_{1,3}^{c} \\
w_{2,1}^{c} & w_{2,2}^{c} & w_{2,3}^{c} \\
w_{3,1}^{c} & w_{3,2}^{c} & w_{3,3}^{c}
\end{bmatrix}
\end{align}
where $c$ stands for channel.

The rule of Sarrus is a well-known method to compute determinant of a 3$\times$3 matrix. As illustrated in Fig. \ref{fig4:sarrus}, when the duplication of the first two columns of the matrix is written along with it, the determinant is calculated by subtracting the sum of the products of three diagonal south-west to north-east lines of elements from the sum of the products of three diagonal north-west to south-east lines of matrix elements. Consequentially, when the kernel size is 3, determinant captures correlation between diagonal components of the depthwise filter by sum of the products, thus determinant is able to capture shape of filter efficiently. 

As the rule of Sarrus, the determinant of $W_{3}^{c}$ is computed as follows:
\begin{align*}
\det W_{3}^{c} = w_{1,1}^{c}w_{2,2}^{c}w_{3,3}^{c} + w_{1,2}^{c}w_{2,3}^{c}w_{3,1}^{c} + w_{1,3}^{c}w_{2,1}^{c}w_{3,2}^{c}\\
- w_{1,3}^{c}w_{2,2}^{c}w_{3,1}^{c} - w_{1,2}^{c}w_{2,1}^{c}w_{3,3}^{c} - w_{1,1}^{c}w_{2,3}^{c}w_{3,2}^{c}
\end{align*}

With the help of determinant, we can shrink the depthwise filter to one dimensional vector $\mathbf{z}\in\mathbb{R}^{C}$.
\vspace{0.2cm}

 \begin{figure}[t]
 \centering
    \includegraphics[scale=0.58]{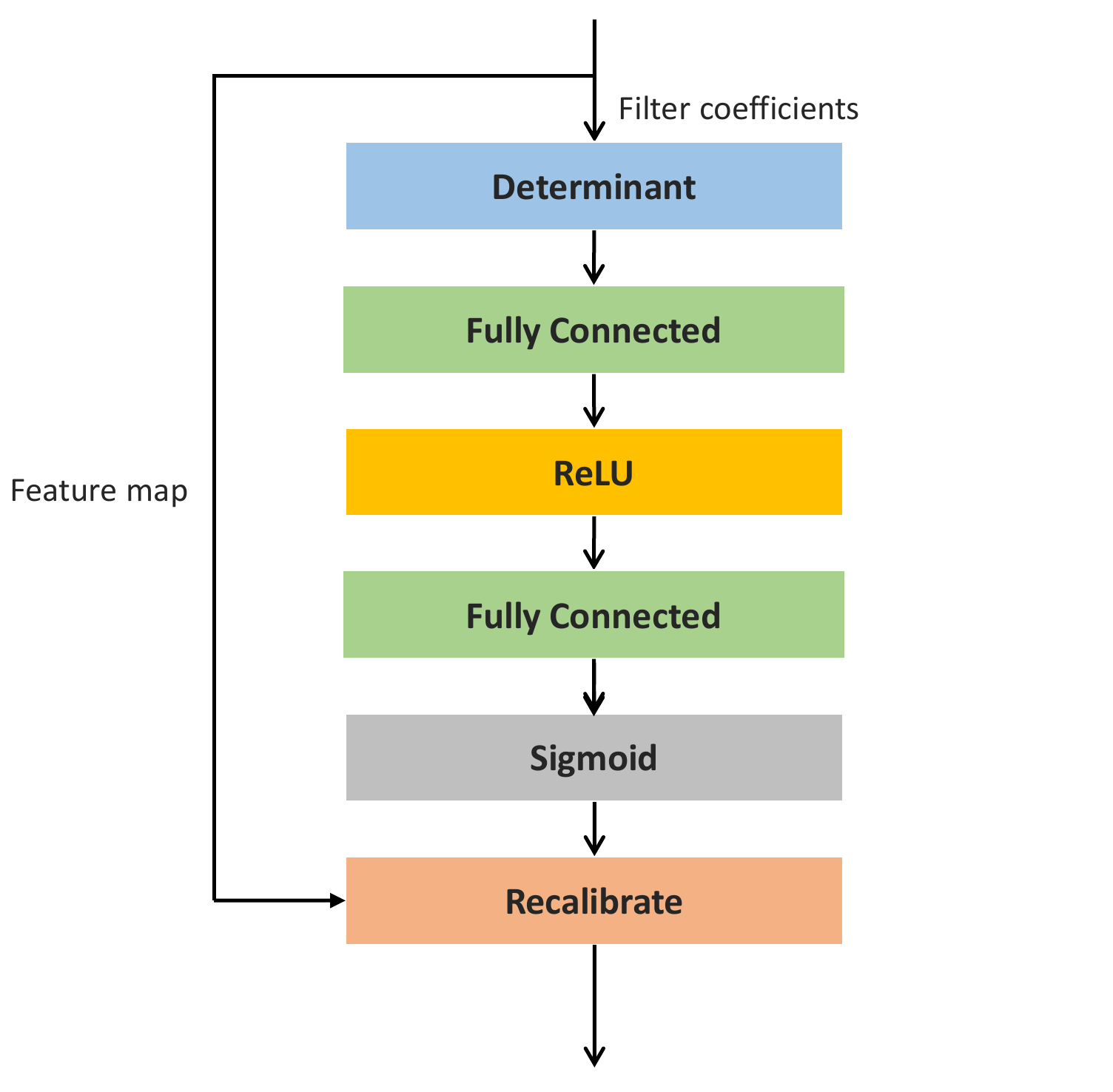}
    \captionsetup{justification=centering}
    \caption{Structure of attention branch.} 
    \label{fig3:attention_branch}
 \end{figure}

 \begin{figure*}[t]
 \centering
    \includegraphics[scale=0.5]{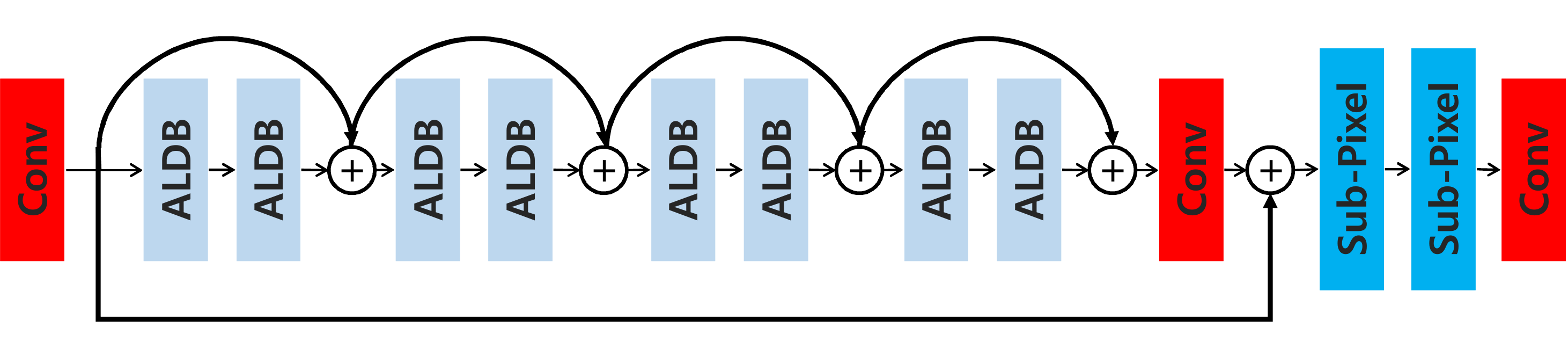}
    \captionsetup{justification=centering}
    \caption{Structure of the proposed ALDSR.} 
    \label{fig5:ALDSR}
 \end{figure*}

\textbf{Attention branch:}              
In contradistinction to most generic CNN attention methods, we utilize interdependency among depthwise filters to get attention. To shrink the depthwise filter information, we take determinant on each depthwise convolution filter to produce one-dimensional vector $\mathbf{z}$ that acts as a depthwise filter descriptor as mentined ealier. To estimate attention across depthwise convolutional layer from the filter description vector $\mathbf{z}$, we opt to employ gating mechanism. As described in \cite{senet}, an adequate gating function can be served by the sigmoid function.
\begin{align}
\mathbf{s} = \sigma (\mathbf{W}_{I}\delta (\mathbf{W}_{D}\mathbf{z}+\mathbf{b}_{D})+\mathbf{b}_{I}),
\end{align}
where $\sigma$ and $\delta$ are sigmoid function and rectified linear unit (ReLU), respectively; $\mathbf{W}_{I}$ and $\mathbf{W}_{D}$ are the weight set of convolutional layers; $\mathbf{b}_{I}$ and $\mathbf{b}_{D}$ are the corresponding biases. To avoid a parameter overhead, the ReLU activation size is set to $\mathbf{z}\in\mathbb{R}^{C/r\times1\times1}$, where $r$ is the reduction ratio. We set the reduction ratio $r$ to 16. Then, we obtain the channel-attention map $\mathbf{s}$ to rescale the feature map $\mathbf{f}$ as follows:
\begin{align}
{\mathbf{f}_{c}}' = s_{c}\cdot\mathbf{f}_{c}
\end{align}

According to \cite{ResidualAttentionNetwork}, stacking attention modules naively leads to hinder the performance by dot production with mask range from zero to one repeatedly. Thus, we counteract this effect by adopting a residual learning strategy \cite{resnet} to make network stable.
\begin{align}
\mathbf{D}_{c} &= \mathbf{f}_{c} + {\mathbf{f}_{c}}' 
\\ &= (1 + s_{c})\cdot\mathbf{f}_{c},
\end{align}
where $\mathbf{D}$ represents the output of attention-aware depthwise convolution.

\section{Attention-Aware Linear Depthwise Network for Image SR}

\subsection{Network Architecture}
In this work, we propose a baseline network for image SR based on the attention-aware linear depthwise block (ALDB), named ALDSR. The overall architecture of ALDSR is illustrated in Fig. \ref{fig5:ALDSR}, which is constructed with the proposed attention-aware linear depthwise block (ALDB). Given $I_{LR}$, ALDSR generates $I_{SR}$ where $I_{LR}$ and $I_{SR}$ stand for low-resolution image and its super-resolved counterpart, respectively. As shown in the figure, two types of residual learning are used to construct the network: (1) global residual learning \cite{VDSR} to provide skip-connection in global scale, and (2) local residual learning to provide skip-connection in every two ALDBs.

According to the former studies \cite{RDN, san, rcan}, we design ALDSR that mainly consists of four parts: shallow feature extraction, deep feature extraction via ALDB, upsample net, and reconstruction part. In ALDSR, only one convolutional layer is used to extract shallow feature $\mathbf{F}_{0}$ from $I_{LR}$ as follows:
\begin{align}
\mathbf{F}_{0} = H_{SF}(\mathbf{I}_{LR}),
\end{align}
where $H_{SF}$ is convolution operation. Then, $\mathbf{F}_{0}$ is used for global residual learning and deep feature extraction with ALDBs as follows:
\begin{align}
\mathbf{F}_{DF} = H_{ALDBs}(\mathbf{F}_{0}),
\end{align}
where $H_{ALDBs}$ is a deep feature extraction structure which consists of several ALDB.
Next, $\mathbf{F}_{DF}$ is upscaled by upsample net as follows:
\begin{align}
\mathbf{F}_{UP} = H_{UP}(\mathbf{F}_{DF}),
\end{align}
where $\mathbf{F}_{UP}$ stands for upsample net. As explored in \cite{EDSR, RDN}, ESPCN \cite{espcn} is used to increase the spatial dimensions of the feature maps. Thus, the upscaled features are passed through one convolutional layer to be mapped into SR image as follows:
\begin{align}
\mathbf{I}_{SR} &= H_{R}(\mathbf{F}_{UP})\\
                &= H_{ALDSR}(\mathbf{I}_{LR}),
\end{align}
where $H_{R}$, $H_{ALDSR}$ denote the reconstruction layer and function of proposed ALDSR, respectively. ALDSR is optimized with a $L_{1}$ loss function which has been demonstrated to be more powerful for performance and convergence \cite{EDSR}.

 \begin{figure}[t]
 \centering
    \includegraphics[scale=0.35]{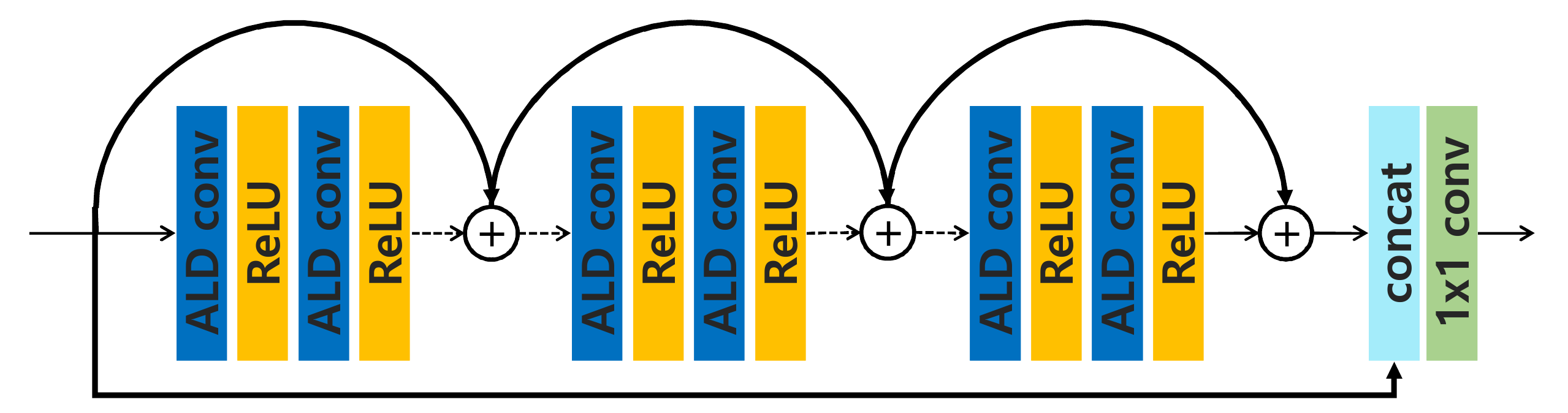}
    \captionsetup{justification=centering}
    \caption{Structure of the attention-aware linear depthwise block (ALDB).} 
    \label{fig6:ALDB}
 \end{figure}

\subsection{Attention-Aware Linear Depthwise Block}
As shown in Fig. \ref{fig5:ALDSR}, ALDSR heavily relies on ALDBs. After applying a $3\times3$ single convolution layer to the input LR images for learning shallow features, a set of ALDBs is employed to learn more discriminative features. We describe the details of the proposed ALDB illustrated in Fig. \ref{fig6:ALDB}. Our ALDB is mainly constructed by cascading several attention-aware linear depthwise separable convolution (ALD convolution) with residual learning. Each ALD convolution consists of attention-aware linear depthwise convolutions followed by $1\times1$ pointwise convolution and ReLU. The pointwise convolution layer is omitted for simplicity in Fig. \ref{fig6:ALDB}.

\begin{table*}[t]
\centering
\caption{ALDSR performance on public benchmark test datasets and DIV2K validation dataset in terms of PSNR(dB) and SSIM).}
\begin{tabular}{|c|cccccc|}
\hline
Dataset          & Bicubic        & VDSR \cite{VDSR}           & LapSRN \cite{lapsrn}         & MemNet \cite{memnet}         & IDN \cite{idn}           & ALDSR   \\ \hline \hline
Set5             & 28.43 / 0.8109 & 31.35 / 0.8838 & 31.54 / 0.8866 & 31.74 / 0.8893 & \textbf{31.82} / \textbf{0.8903} & 31.78 / 0.8895 \\ 
Set14            & 26.00 / 0.7023 & 28.02 / 0.7678 & 28.09 / 0.7694 & 28.26 / 0.7723 & 28.25 / 0.7730 & \textbf{28.37} / \textbf{0.7760} \\ 
B100             & 25.96 / 0.6678 & 27.29 / 0.7252 & 27.32 / 0.7264 & 27.40 / 0.7281 & \textbf{27.41} / 0.7297 & \textbf{27.41} / \textbf{0.7304} \\ 
Urban100         & 23.14 / 0.6574 & 25.18 / 0.7525 & 25.21 / 0.7553 & 25.50 / 0.7630 & 25.41 / 0.7632 & \textbf{25.55} / \textbf{0.7685} \\
DIV2K validation & 28.11 / 0.775  & 29.82 / 0.824  & 29.88 / 0.825  & -              & -              & \textbf{30.16} / \textbf{0.8313} \\ \hline
Parameters       & -              & 665k           & 812k           & 677k           & 796k           & 731k           \\ \hline
\end{tabular}
\label{table0:aldsr_benchmark_results}
\end{table*}

\begin{table*}[t]
\centering
\caption{ALD-RDN performance on public benchmark test datasets and DIV2K validation dataset in terms of PSNR(dB) and SSIM.}
\begin{tabular}{|c|cc|cc|}
\hline
Dataset   & RDN \cite{RDN}     & RDN(re-implemented)    & DW-RDN   & ALD-RDN                                                    \\ \hline \hline
Set5  & \begin{tabular}[c]{@{}c@{}}32.47   / 0.8990\end{tabular} & \begin{tabular}[c]{@{}c@{}}32.40  / 0.8977\end{tabular} & 32.10 / 0.8937   & \begin{tabular}[c]{@{}c@{}}\textbf{32.18}   / \textbf{0.8954}\end{tabular} \\ 
Set14                                                        & \begin{tabular}[c]{@{}c@{}}28.81   / 0.7871\end{tabular} & \begin{tabular}[c]{@{}c@{}}28.73   / 0.7861\end{tabular} & 28.59 / 0.7812                                             & \begin{tabular}[c]{@{}c@{}}\textbf{28.62}   / \textbf{0.7837}\end{tabular} \\ 
B100                                                         & \begin{tabular}[c]{@{}c@{}}27.72   / 0.7419\end{tabular} & \begin{tabular}[c]{@{}c@{}}27.68   / 0.7397\end{tabular} & 27.57 / 0.7354                                             & \begin{tabular}[c]{@{}c@{}}\textbf{27.61}   / \textbf{0.7378}\end{tabular} \\ 
Urban100                                                     & \begin{tabular}[c]{@{}c@{}}26.61   / 0.8028\end{tabular} & \begin{tabular}[c]{@{}c@{}}26.49   / 0.7988\end{tabular} & 26.06 / 0.7843                 & \begin{tabular}[c]{@{}c@{}}\textbf{26.23}   / \textbf{0.7914}\end{tabular} \\ 
\begin{tabular}[c]{@{}c@{}}DIV2K   validation\end{tabular} & -                                                          & \begin{tabular}[c]{@{}c@{}}30.65   / 0.8430\end{tabular} & \begin{tabular}[c]{@{}c@{}}30.45   / 0.8375\end{tabular} & \begin{tabular}[c]{@{}c@{}}\textbf{30.54}   / \textbf{0.8406}\end{tabular} \\ \hline
\end{tabular}
\label{table1:benchmark_results}
\end{table*}

Similar to the ALDSR architecture, local residual learning is also employed in ALDB by adding short-path skip-connection in every two ALD convolutions. Furthermore, we adopt contiguous memory mechanism \cite{RDN} which reads state from the preceding building block to make full use of hierarchical features that is able to get more information for SR reconstruction. However, it puts high computational burden on the network, thus it is required to simplify it. The modified contiguous memory mechanism is realized by passing the state of the preceding ALDB only to $1\times1$ bottleneck layer of current ALDB. Then, it is concatenated with the final residual output of current ALDB. Bottleneck layer which uses $1\times1$ convolution is used to adaptively fuse the concatenated features and reduce the channel size to half. 

\section{Experiments}
\subsection{Experimental Setup}
The DIV2K \cite{div2k} dataset consists of 800 high-resolution images for training, 100 images for validation, and 100 images for test. Since the ground truth in the test images is not released, we compare performance on the 100 validation images. We also use four standard benchmark datasets: Set5 \cite{set5}, Set14 \cite{set14}, BSD100 \cite{b100}, and Urban100 \cite{urban100}. For experiments, we generate LR images by 4$\times$ bicubic-downsampling from HR images. We evaluate SR results in Y channel of the transformed YCbCr space in terms of peak signal-to-noise ratio (PSNR) and structural similarity (SSIM). For training, 16 LR color patches of size 48$\times$48 from LR images with the corresponding HR patches provided as inputs. We augment the patches by randomly flipping horizontally or vertically and rotating $90^{\circ}$. We adopt Adam Optimizer \cite{adam} for training with the initial learning rate of $10^{-4}$ and halved once at epoch 200. We train all models by 300 epochs. All experiments are implemented on PyTorch framework \cite{pytorch}.

\subsection{Performance Evaluation and Analysis}
\textbf{ALD:}
We first compare the proposed ALDSR with other state-of-the-art SR methods including VDSR \cite{VDSR}, LapSRN \cite{lapsrn}, MemNet \cite{memnet}, IDN \cite{idn}. Table \ref{table0:aldsr_benchmark_results} shows the average PSNR and SSIM values on four benchmark datasets and DIV2K validation dataset. Our ALDSR exhibits a significant improvement compared to the other methods on most datasets. The gaps is obviously noticeable on Set14 with outperforming MemNet \cite{memnet} by 0.11dB.

To further investigate the effectiveness of the proposed ALDNet, we apply attention-aware linear depthwise convolution to a state-of-the-art architecture, Residual Dense Network \cite{RDN}, which is constructed to cascade several building blocks so-called RDB (Residual Dense Block). We construct ALDNet equivalent of RDN by simply replacing standard convolutional layers in RDB to ALD convolutional layers (Fig. \ref{fig2:ALD}). Furthermore, we also construct depthwise(DW)-RDN where standard convolutional layers are replaced with depthwise separable convolutional layers (Fig. \ref{fig1_a:DW}) in RDB.

The benchmark results of ALD-RDN are reported in Table \ref{table1:benchmark_results}, which shows the overall average PSNR and SSIM. As shown in Table \ref{table1:benchmark_results}, compared with DW-RDN, our ALD-RDN produces better results for all benchmark datasets. Especially, on the Urban100 dataset which contains images with complex patterns, ALD-RDN shows significant performance improvement, which indicates ALD-RDN is very effective for image SR with complex patterns. Since RDN requires quite higher computational complexity than DW-RDN and ALD-RDN, the original RDN achieves the best performance. Also, since the original RDN is trained over 300 epochs, it obtains better results than reimplemented version. The number of parameters of building blocks among them is reported in Table \ref{table4:parameters}.
\vspace{0.2cm}

\begin{table}[t]
\centering
\caption{Comparison among building blocks in terms of the number of parameters.}
\begin{tabular}{|c|c|}
\hline
\begin{tabular}[c]{@{}c@{}}Building block\end{tabular} & \begin{tabular}[c]{@{}c@{}}Number of parameters\end{tabular} \\ \hline \hline
RDB        & 1,363,968 \\ 
DW-RDB     & 205,056   \\ 
LDW-RDB    & 205,056   \\ 
ALD-RDB    & 257,280   \\ 
ALDB       & 41,472    \\ \hline
\end{tabular}
\label{table4:parameters}
\end{table}

\begin{table*}[t]
\centering
\caption{Performance comparison by using different descriptors.}
\begin{tabular}{|c|ccc|ccc|}
\hline
\multirow{2}{*}{Dataset} & \multicolumn{3}{c|}{ALDSR}                       & \multicolumn{3}{c|}{ALD-RDN}                                                                 \\ \cline{2-7} 
                         & Average        & Max            & Determinant    & Average        & Max            & Determinant                                                \\ \hline \hline
Set5                     & 31.77 / 0.8892 & 31.76 / 0.8892 & \textbf{31.78} / \textbf{0.8895} & 32.20 / 0.8953 & \textbf{32.23} / \textbf{0.8957} & 32.18 / 0.8954                                             \\ 
Set14                    & \textbf{28.37} / 0.7758 & 28.36 / 0.7757 & \textbf{28.37} / \textbf{0.7760} & \textbf{28.66} / \textbf{0.7837} & 28.63 / 0.7827 & 28.62 / \textbf{0.7837}
               \\ 
B100                     & \textbf{27.41} / 0.7302 & \textbf{27.41} / 0.7302 & \textbf{27.41} / \textbf{0.7304} & 27.60 / 0.7375 & \textbf{27.61} / 0.7372 & \textbf{27.61} / \textbf{0.7378}                                             \\ 
Urban100                 & \textbf{25.56} / 0.7680 & \textbf{25.56} / 0.7680 & 25.55 / \textbf{0.7685} & 26.20 / 0.7905 & \textbf{26.23} / 0.7908 & \textbf{26.23} / \textbf{0.7914}                                             \\ 
DIV2K validation         & \textbf{30.17} / \textbf{0.8313} & 30.16 / 0.8311 & 30.16 / \textbf{0.8313} & 30.52 / 0.8402 & \textbf{30.54} / 0.8400 & \textbf{30.54} / \textbf{0.8406}                                             \\ \hline
\end{tabular}
\label{table3:different_descriptor}
\end{table*}

\begin{table}[t]
\centering
\caption{Performance comparison between depthwise convolution and linear depthwise convolution.}
\begin{tabular}{|c|cc|}
\hline
Dataset     & DW-RDN            & LDW-RDN        \\ \hline \hline
Set5 & \begin{tabular}[c]{@{}c@{}}32.10 / 0.8937 \end{tabular} &     \begin{tabular}[c]{@{}c@{}}\textbf{32.17} / \textbf{0.8950}\end{tabular} \\ 
Set14       & 28.59 / 0.7812    & \textbf{28.60} / \textbf{0.7826}    \\ 
B100        & 27.57 / 0.7354    & \textbf{27.59} / \textbf{0.7367}    \\ 
Urban100    & 26.06 / 0.7843    & \textbf{26.14} / \textbf{0.7883}    \\ 
\begin{tabular}[c]{@{}c@{}}DIV2K validation\end{tabular} & \begin{tabular}[c]{@{}c@{}}30.45   / 0.8375\end{tabular} & \begin{tabular}[c]{@{}c@{}}\textbf{30.50}   / \textbf{0.8393}\end{tabular} \\ \hline
\end{tabular}
\label{table2:linearity}
\end{table}
\textbf{Effect of linearity:}
To verify whether the linear depthwise convolution performs a crucial role in performance, we construct LDW-RDN. Specifically, we remove "ReLU" between depthwise convolution and pointwise convolution in DW-RDB, and the corresponding results are reported in Table \ref{table2:linearity}. We observe that linear depthwise convolution improves performance remarkably without any additional parameters. Especially, LDW-RDN yields a performance improvement from 26.06dB to 26.14dB on the Urban100 dataset which contains complex images and is difficult to reconstructed. Thus, linear depthwise convolution prevents destruction of informative features to reconstruct SR image from non-linearity such as "ReLU", and thus it performs better on the images with complex structures.
\vspace{0.2cm}

\textbf{Effect of determinant descriptor:}
We also examine the effect of the determinant descriptor. To that end, we select average and max pooling to describe depthwise filter characteristics and conduct experiments with ALDSR and ALD-RDN. The results are reported in Table \ref{table3:different_descriptor}. Although all descriptors perform well on ALDSR, the average descriptor achieves slightly better PSNR performance than the others in most datasets. On the other hand, on ALD-RDN, max pooling generally produces better PSNR results. Determinant descriptor, which is used for ALDNet, exhibits the best SSIM results in all benchmark datasets except for Set5 on ALD-RDN, as well as it obtains comparable PSNR results to the others.
\vspace{0.2cm}

\textbf{Discussion:}
Unlike standard convolution, depthwise convolution needs very few parameters. Since the maximum value of them represents a value with the most dominant influence on the convolution transform, it is effective to describe the convolution which has few parameters by max pooling. Thus, max pooling shows better PSNR results. However, max pooling describes only one value in the depthwise filter, and thus it is difficult to capture any information on describing the entire parameters of filter by max pooling. For this reason, max pooling shows unfavorable SSIM results. On the other hand, since determinant considers all the parameters carefully in the depthwise filter enabling network to learn importance of depthwise filter based on a lot of information of filter including shape, it can be found that not only the PSNR is high but also the SSIM is significantly higher.
\vspace{0.2cm}

\begin{figure*}[t]
\centering

\begin{tabular}{cccc}
\begin{adjustbox}{valign=t}
\begin{tabular}{@{}c@{}}
  \includegraphics[scale=0.40]{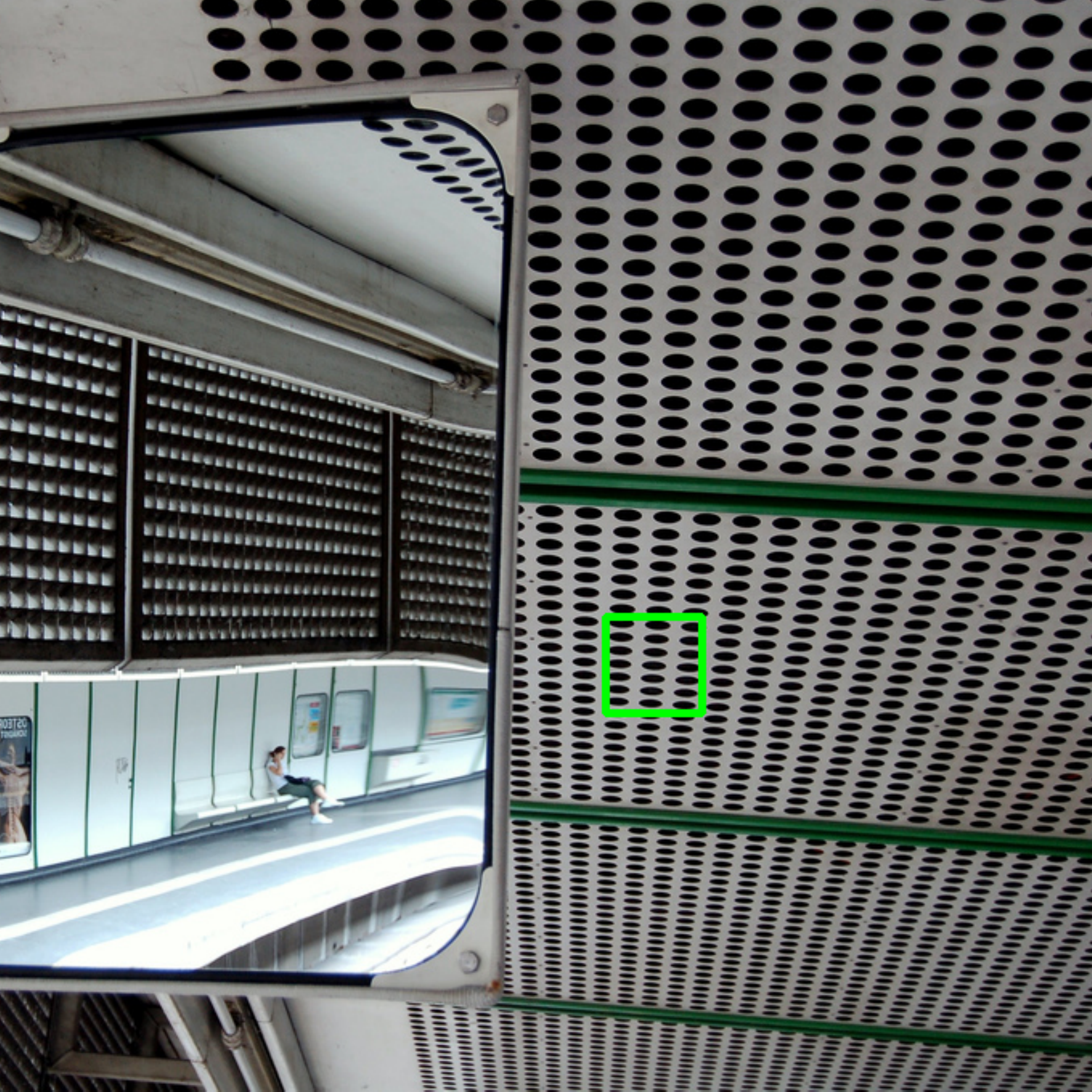}\\ 
  img004 from Urban100 \\
\end{tabular}
\end{adjustbox}

&
\begin{adjustbox}{valign=t}
\begin{tabular}{@{}c@{}}
  \includegraphics[scale=1.95]{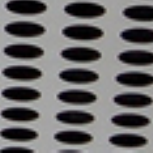} \\ 
  \begin{tabular}[c]{@{}c@{}}HR\\ (PSNR/SSIM)\end{tabular}\\ 
  \includegraphics[scale=1.95]{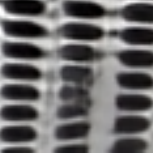} \\
  \begin{tabular}[c]{@{}c@{}}LapSRN\\ (22.41dB/0.7989)\end{tabular} \\
\end{tabular}
\end{adjustbox}

&
\begin{adjustbox}{valign=t}
\begin{tabular}{@{}c@{}}
  \includegraphics[scale=1.95]{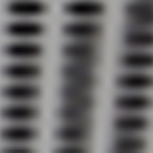} \\ 
  \begin{tabular}[c]{@{}c@{}}Bicubic\\ (21.09dB/0.6792)\end{tabular} \\ 
  \includegraphics[scale=1.95]{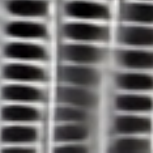} \\ 
  \begin{tabular}[c]{@{}c@{}}IDN\\ (22.27dB/0.7972)\end{tabular} \\ 
\end{tabular}
\end{adjustbox}

&
\begin{adjustbox}{valign=t}
\begin{tabular}{@{}c@{}}
  \includegraphics[scale=1.95]{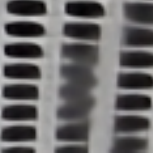} \\ 
  \begin{tabular}[c]{@{}c@{}}VDSR\\ (22.42dB/0.7955)\end{tabular} \\ 
  \includegraphics[scale=1.95]{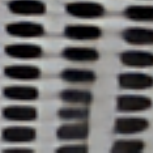} \\ 
  \begin{tabular}[c]{@{}c@{}}ALDSR\\ (22.95dB/0.8201)\end{tabular} \\ 
\end{tabular}
\end{adjustbox}

\end{tabular}
%

\begin{tabular}{cccc}
\\
\begin{adjustbox}{valign=t}
\begin{tabular}{@{}c@{}}
  \includegraphics[scale=0.40]{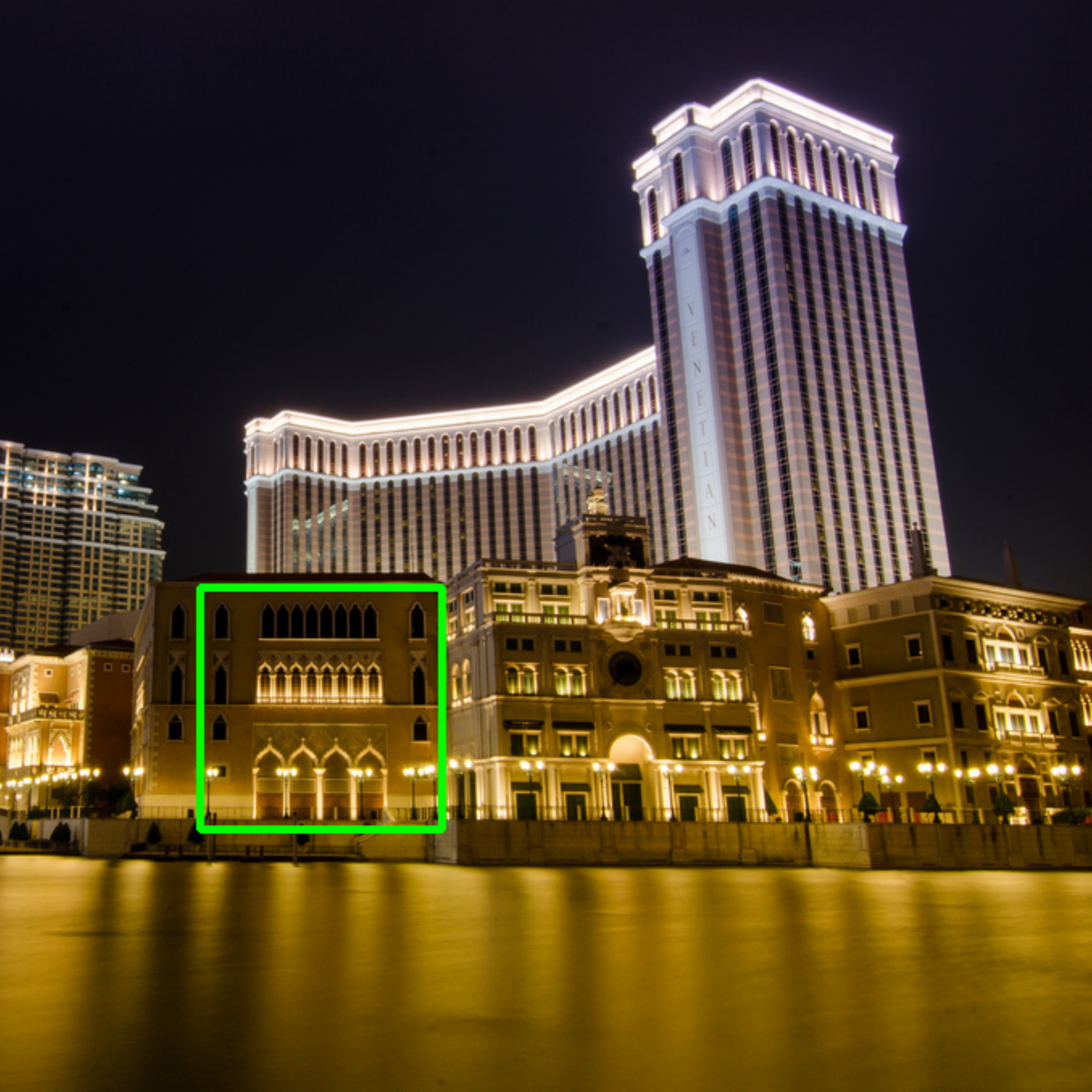}\\ 
  img085 from Urban100 \\
\end{tabular}
\end{adjustbox}

&
\begin{adjustbox}{valign=t}
\begin{tabular}{@{}c@{}}
  \includegraphics[scale=0.78]{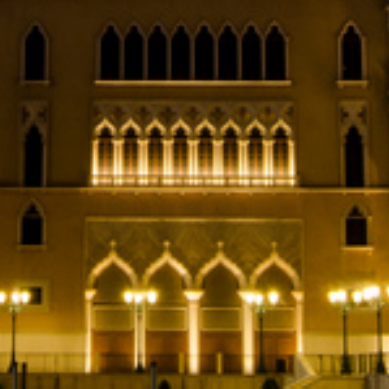} \\ 
  \begin{tabular}[c]{@{}c@{}}HR\\ (PSNR/SSIM)\end{tabular}\\ 
  \includegraphics[scale=0.78]{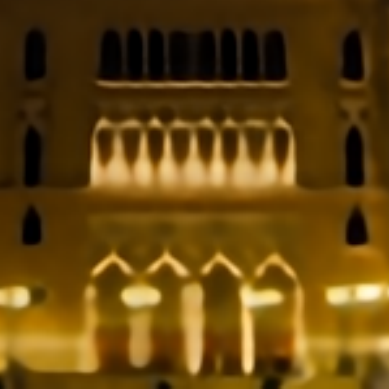} \\
  \begin{tabular}[c]{@{}c@{}}LapSRN\\ (27.10dB/0.8785)\end{tabular} \\
\end{tabular}
\end{adjustbox}

&
\begin{adjustbox}{valign=t}
\begin{tabular}{@{}c@{}}
  \includegraphics[scale=0.78]{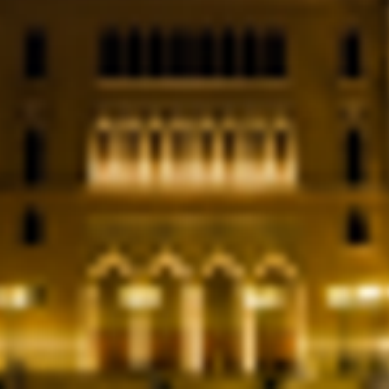} \\ 
  \begin{tabular}[c]{@{}c@{}}Bicubic\\ (25.90dB/0.8371)\end{tabular} \\ 
  \includegraphics[scale=0.78]{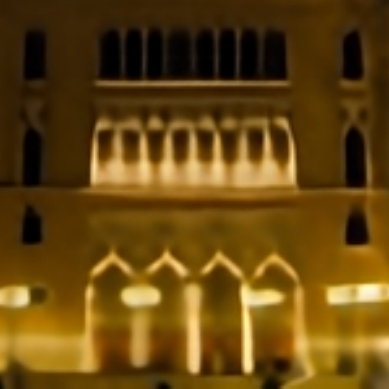} \\ 
  \begin{tabular}[c]{@{}c@{}}IDN\\ (27.13dB/0.8796)\end{tabular} \\ 
\end{tabular}
\end{adjustbox}

&
\begin{adjustbox}{valign=t}
\begin{tabular}{@{}c@{}}
  \includegraphics[scale=0.78]{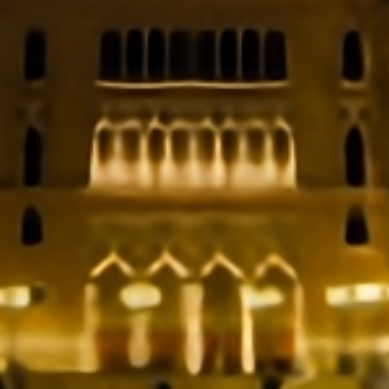} \\ 
  \begin{tabular}[c]{@{}c@{}}VDSR\\ (27.14dB/0.8773)\end{tabular} \\ 
  \includegraphics[scale=0.78]{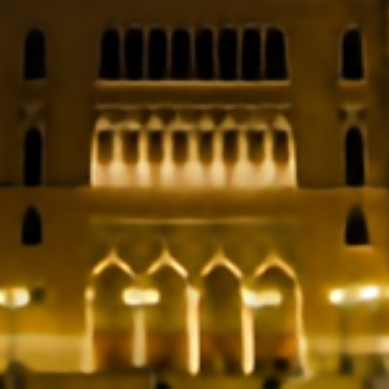} \\ 
  \begin{tabular}[c]{@{}c@{}}ALDSR\\ (27.75dB/0.8906)\end{tabular} \\ 
\end{tabular}
\end{adjustbox}

\end{tabular}
\caption{Visual comparison for 4$\times$ SR on Urban100 dataset.}
\label{fig4:aldsr_visual_comparison}
\end{figure*}

\begin{figure*}[t]

\begin{tabular}{ccccc}
\begin{adjustbox}{valign=t}
\begin{tabular}{@{}c@{}}
  \includegraphics[scale=0.38]{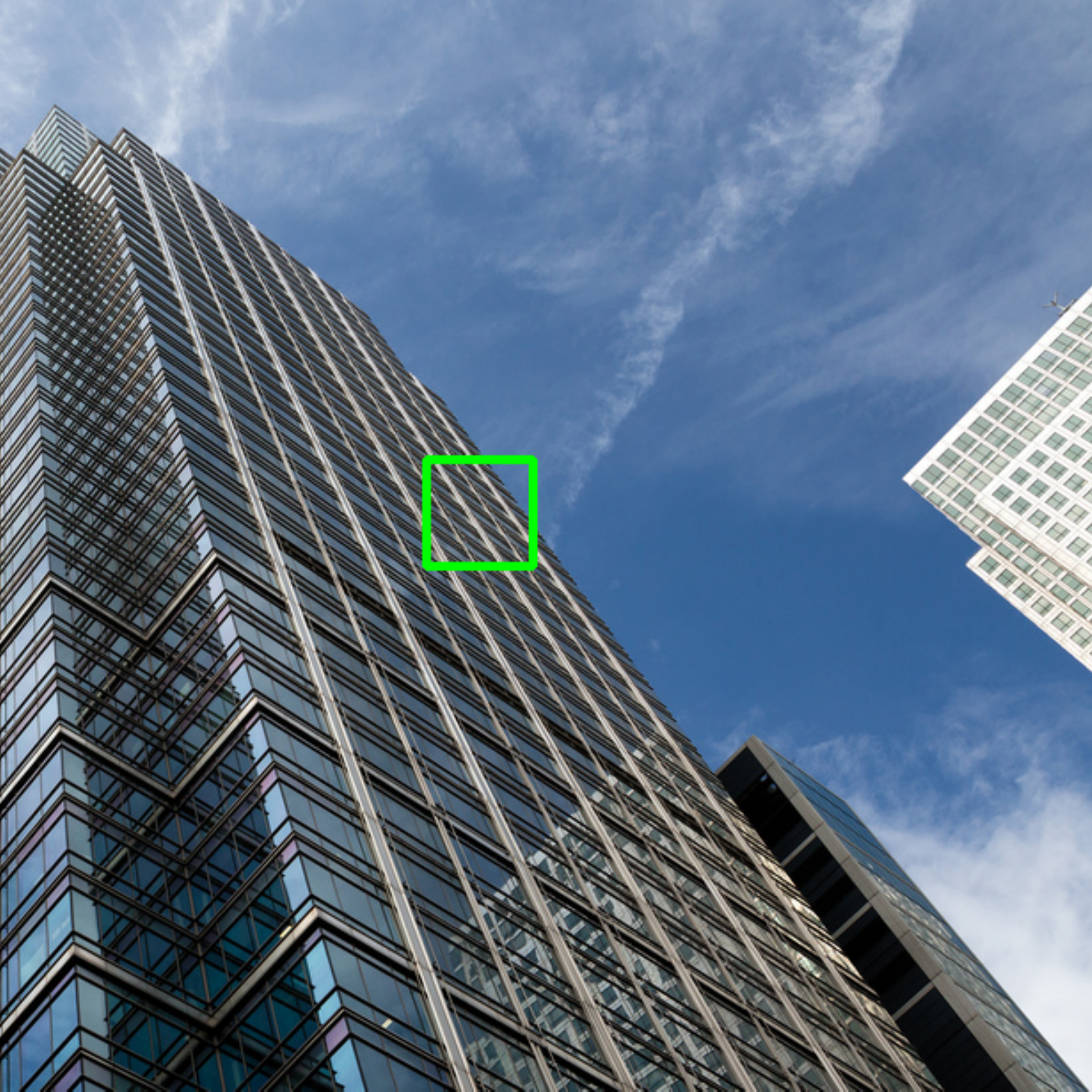}\\ 
  img047 from Urban100 \\
\end{tabular}
\end{adjustbox}

&
\begin{adjustbox}{valign=t}
\begin{tabular}{@{}c@{}}
  \includegraphics[scale=1.6]{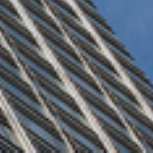} \\ 
  \begin{tabular}[c]{@{}c@{}}HR\\ (PSNR/SSIM)\end{tabular}\\ 
  \includegraphics[scale=1.6]{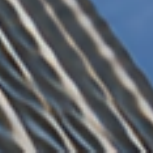} \\
  \begin{tabular}[c]{@{}c@{}}LDW-RDN\\ (21.94dB/0.7713)\end{tabular} \\
\end{tabular}
\end{adjustbox}

&
\begin{adjustbox}{valign=t}
\begin{tabular}{@{}c@{}}
  \includegraphics[scale=1.6]{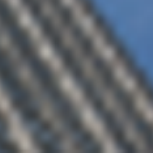} \\ 
  \begin{tabular}[c]{@{}c@{}}Bicubic\\ (20.01dB/0.6282)\end{tabular} \\ 
  \includegraphics[scale=1.6]{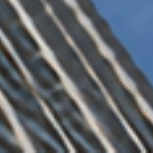} \\ 
  \begin{tabular}[c]{@{}c@{}}ALD-RDN(avr)\\ (21.96dB/0.7747)\end{tabular} \\ 
\end{tabular}
\end{adjustbox}

&
\begin{adjustbox}{valign=t}
\begin{tabular}{@{}c@{}}
  \includegraphics[scale=1.6]{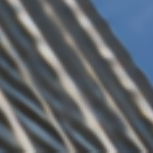} \\ 
  \begin{tabular}[c]{@{}c@{}}RDN(re-im)\\ (22.16dB/0.7827)\end{tabular} \\ 
  \includegraphics[scale=1.6]{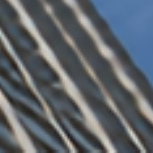} \\ 
  \begin{tabular}[c]{@{}c@{}}ALD-RDN(max)\\ (21.94dB/0.7744)\end{tabular} \\ 
\end{tabular}
\end{adjustbox}

&
\begin{adjustbox}{valign=t}
\begin{tabular}{@{}c@{}}
  \includegraphics[scale=1.6]{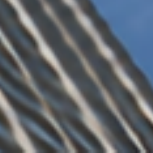} \\ 
  \begin{tabular}[c]{@{}c@{}}DW-RDN\\ (21.90dB/0.7653)\end{tabular} \\ 
  \includegraphics[scale=1.6]{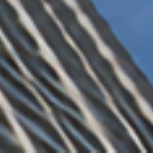} \\
  \begin{tabular}[c]{@{}c@{}}ALD-RDN(det)\\ (22.00dB/0.7778)\end{tabular} \\ 
\end{tabular}
\end{adjustbox}
\end{tabular}
%

\begin{tabular}{ccccc}
\\
\begin{adjustbox}{valign=t}
\begin{tabular}{@{}c@{}}
  \includegraphics[scale=0.32]{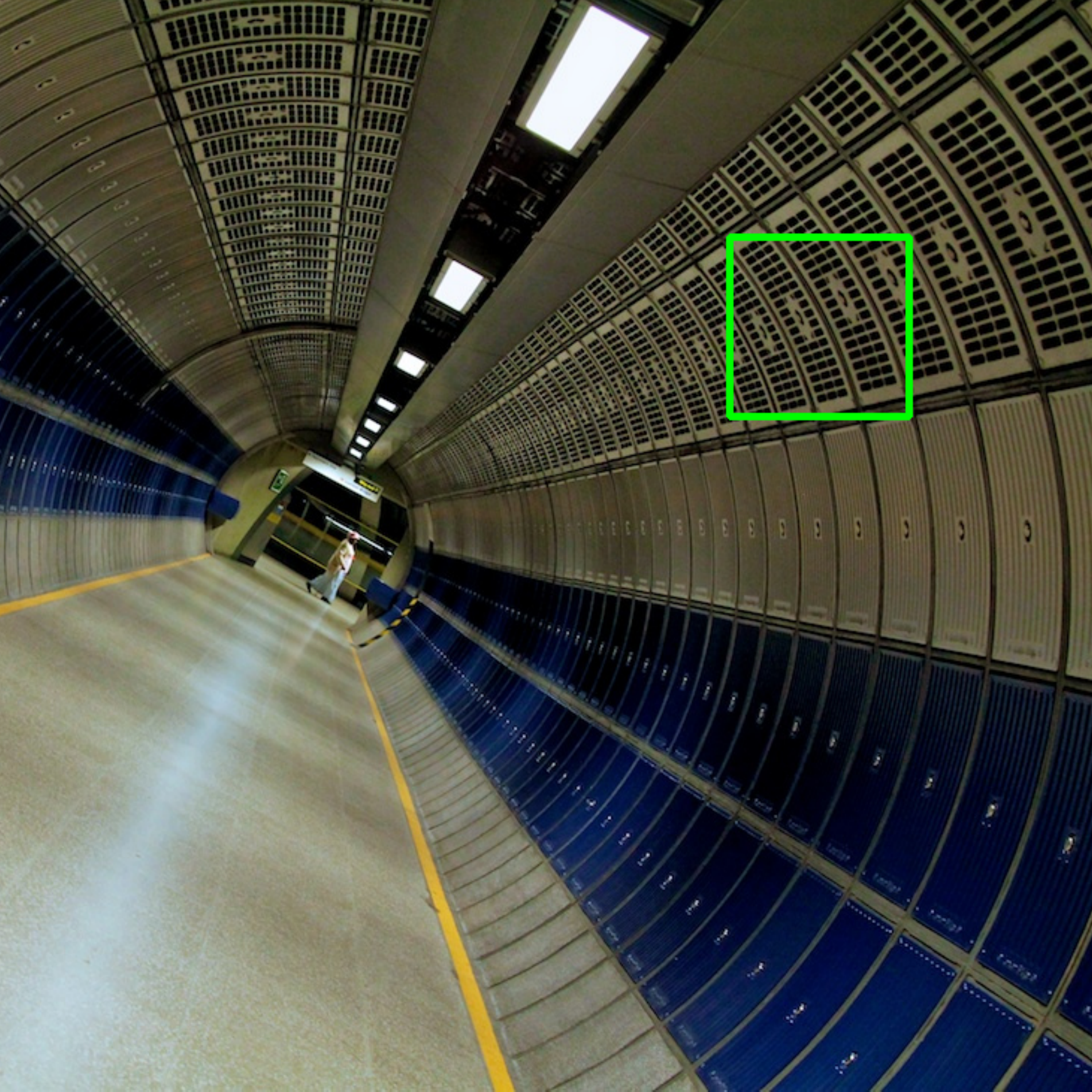}\\ 
  img078 from Urban100 \\
\end{tabular}
\end{adjustbox}

&
\begin{adjustbox}{valign=t}
\begin{tabular}{@{}c@{}}
  \includegraphics[scale=0.8]{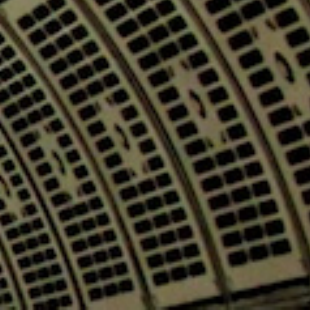} \\ 
  \begin{tabular}[c]{@{}c@{}}HR\\ (PSNR/SSIM)\end{tabular}\\ 
  \includegraphics[scale=0.8]{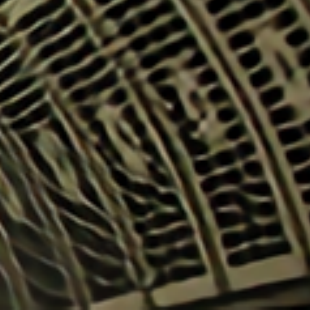} \\
  \begin{tabular}[c]{@{}c@{}}LDW-RDN\\ (27.38dB/0.7791)\end{tabular} \\
\end{tabular}
\end{adjustbox}

&
\begin{adjustbox}{valign=t}
\begin{tabular}{@{}c@{}}
  \includegraphics[scale=0.8]{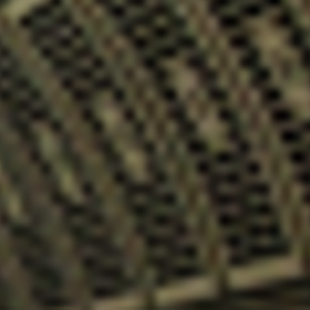} \\ 
  \begin{tabular}[c]{@{}c@{}}Bicubic\\ (25.71dB/0.6797)\end{tabular} \\ 
  \includegraphics[scale=0.8]{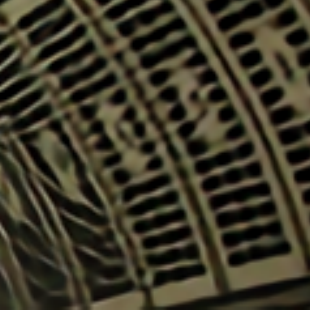} \\ 
  \begin{tabular}[c]{@{}c@{}}ALD-RDN(avr)\\ (27.59dB/0.7825)\end{tabular} \\ 
\end{tabular}
\end{adjustbox}

&
\begin{adjustbox}{valign=t}
\begin{tabular}{@{}c@{}}
  \includegraphics[scale=0.8]{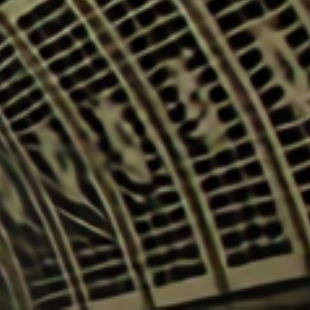} \\ 
  \begin{tabular}[c]{@{}c@{}}RDN(re-im)\\ (27.93dB/0.7917)\end{tabular} \\ 
  \includegraphics[scale=0.8]{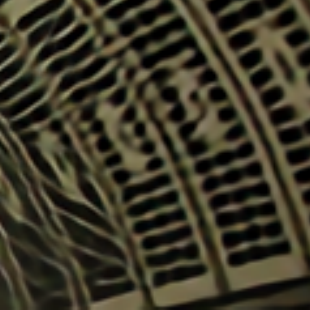} \\ 
  \begin{tabular}[c]{@{}c@{}}ALD-RDN(max)\\ (27.52dB/0.7807)\end{tabular} \\ 
\end{tabular}
\end{adjustbox}

&
\begin{adjustbox}{valign=t}
\begin{tabular}{@{}c@{}}
  \includegraphics[scale=0.8]{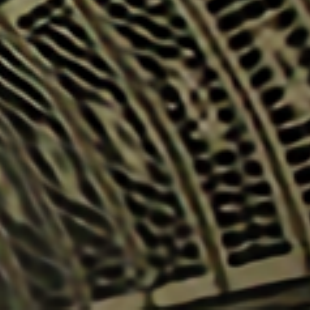} \\ 
  \begin{tabular}[c]{@{}c@{}}DW-RDN\\ (27.10dB/0.7706)\end{tabular} \\ 
  \includegraphics[scale=0.8]{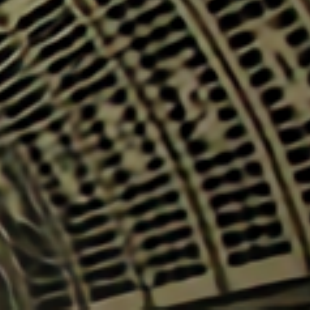} \\
  \begin{tabular}[c]{@{}c@{}}ALD-RDN(det)\\ (27.70dB/0.7842)\end{tabular} \\ 
\end{tabular}
\end{adjustbox}

\end{tabular}
\caption{Visual comparison for 4$\times$ SR on Urban100 dataset.}
\label{fig5:aldrdn_visual_comparison}
\end{figure*}

\textbf{Visual Quality:}
We provide visual comparison among ALDSR and state-of-the-art SR methods in Fig. \ref{fig4:aldsr_visual_comparison}. For challenging details in images "img004" from Urban100,  most SR methods suffer from undesirable artifacts and heavy blurry results, however the proposed ALDSR generates good SR image. In addition, severe distortions are found in some reconstruction results by other SR methods, whereas ALDSR can reconstruct the repetitive patterns well without severe distortions. We also provide visual comparison among RDN, DW-RDN, LDW-RDN and ALD-RDN in Fig. \ref{fig5:aldrdn_visual_comparison}. It can be observed that the proposed ALD-RDN exhibits comparable visual quality to RDN and even achieves better performance in lattice. In "img078" from Urban100, bicubic interpolation loses details and texture resulting in a very blurry image. DW-RDN is able to recover edges and coarse details, but it still fails to get fine details with wrong lattice. LDW-RDN produces better reconstruction results than DW-RDN. Compared with the ground-truth, our ALD-RDN reconstructs more convincing SR images with accurate lattice and fine details even obtaining better visual quality than RDN. It can be observed that determinant descriptor shows superior performance to average and max descriptors. These observations ensure the effectiveness of determinant-based ALDNet with powerful representation ability and reconstruction performance by informative feature preservation.

\section{Conclusion}
In this paper, we have proposed an attention-aware linear depthwise network (ALDNet) for image SR that preserves informative features and enhances representation ability while reducing computational burden. Linear depthwise convolution allows ALDNet to prevent destruction of useful information that provides clues for SR reconstruction. Moreover, attention branch on linear depthwise convolution layer learns importance of each depthwise convolution filter, and thus allows ALDNet to strengthen informative features. To facilitate depthwise convolution, we use determinant as a linear depthwise convolution descriptor because the depthwise filter can be represented by square matrix. Experimental results demonstrate that ALDNet achieves state-of-the-art SR performance with low computational cost.  

\ifCLASSOPTIONcaptionsoff
  \newpage
\fi

\bibliographystyle{IEEETran}
\bibliography{bibliography}




\begin{IEEEbiography}[{\includegraphics[width=1in,height=1.25in,clip,keepaspectratio]{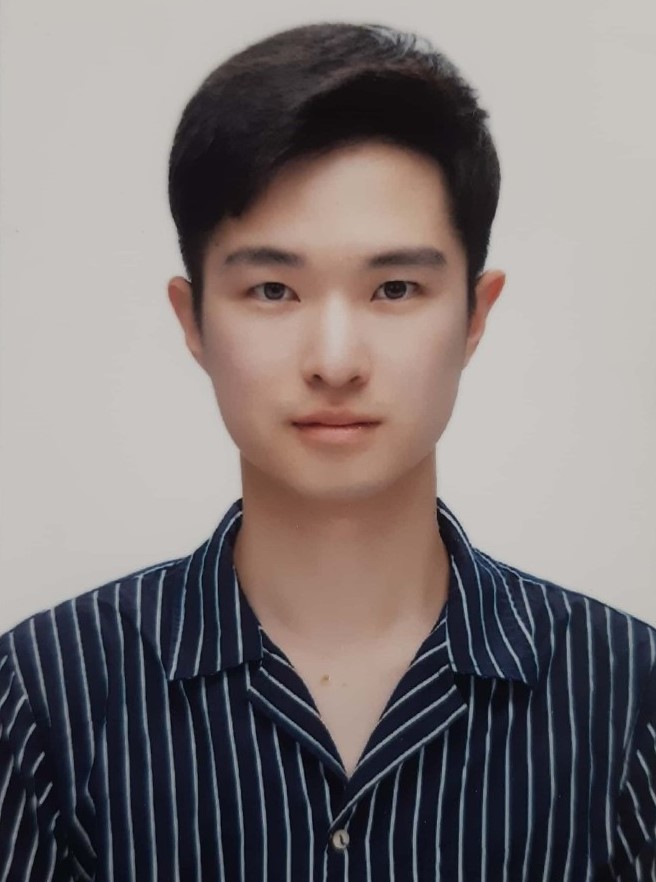}}]{Seongmin Hwang} received the B.S. and M.S. degrees in electronic engineering from Chonnam National University, South Korea, in 2017 and 2019, respectively. He is currently pursing the Ph.D. degree in the same university. His research interests include image processing, computer vision, machine learning and deep learning.
\end{IEEEbiography}

\begin{IEEEbiography}[{\includegraphics[width=1in,height=1.25in,clip,keepaspectratio]{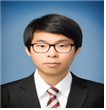}}]{Gwanghuyn Yu} received the B.S. degree in electronic engineering from Chosun University, South Korea in 2016. He got the M.S. degree in electronic engineering from Chonnam National University, South Korea in 2018. Since 2018, he has been a Ph.D. student in the same university. Also, he is a CEO from Insectpedia company, South Korea. His main research area includes digital signal processing, image processing, speech signal processing and machine learning.
\end{IEEEbiography}

\begin{IEEEbiography}[{\includegraphics[width=1in,height=1.25in,clip,keepaspectratio]{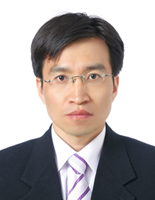}}]{Cheolkon Jung} (M'08) received the B.S., M.S., and Ph.D. degrees in electronic engineering from Sungkyunkwan University, South Korea, in 1995, 1997, and 2002, respectively. He was a Research Staff Member with the Samsung Advanced Institute of Technology (Samsung Electronics), South Korea, from 2002 to 2007. He was a Research Professor with the School of Information and Communication Engineering, Sungkyunkwan University, from 2007 to 2009. Since 2009, he has been with the School of Electronic Engineering, Xidian University, China, where he is currently a Full Professor and the Director of the Xidian Media Lab. His main research interests include image and video processing, computer vision, pattern recognition, machine learning, computational photography, video coding, virtual reality, information fusion, multimedia content analysis and management, and 3DTV.
\end{IEEEbiography}

\begin{IEEEbiography}[{\includegraphics[width=1in,height=1.25in,clip,keepaspectratio]{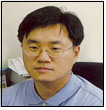}}]{Jinyong Kim} received the B.S., M.S., and Ph.D. degrees in electronic engineering from Seoul National University, South Korea, in 1986, 1988, and 1994, respectively. He was a full time researcher at Korea Telecom Software Research Center, South Korea, from 1993 to 1994. Since 1995, he has been a Full Professor with School of Electronics and Computer Engineering, Chonnam National University, South Korea. His main research topics include digital signal processing, image processing, speech signal processing and machine learning.
\end{IEEEbiography}




\end{document}